\documentclass[prd,superscriptaddress,showpacs,nofootinbib,amsmath,amssymb,aps,11pt]{revtex4}

\usepackage{bm}
\usepackage{amsfonts}
\usepackage{latexsym}
\usepackage[latin1]{inputenc}
\usepackage{graphicx}
\usepackage{amsmath}
\usepackage{palatino}
\usepackage{mathpazo}
\usepackage{booktabs}
\usepackage{dcolumn}

\def\jnl@style{\it}
\def\aaref@jnl#1{{\jnl@style#1}}

\def\aaref@jnl#1{{\jnl@style#1}}

\def\aj{\aaref@jnl{AJ}}                   
\def\apj{\aaref@jnl{ApJ}}                 
\def\apjl{\aaref@jnl{ApJ}}                
\def\apjs{\aaref@jnl{ApJS}}               
\def\apss{\aaref@jnl{Ap\&SS}}             
\def\aap{\aaref@jnl{A\&A}}                
\def\aapr{\aaref@jnl{A\&A~Rev.}}          
\def\aaps{\aaref@jnl{A\&AS}}              
\def\mnras{\aaref@jnl{Mon.~Not.~Roy.~Astron.~Soc.}}             
\def\prd{\aaref@jnl{Phys.~Rev.~D}}        
\def\prc{\aaref@jnl{Phys.~Rev.~C}}  
\def\prl{\aaref@jnl{Phys.~Rev.~Lett.}}    
\def\qjras{\aaref@jnl{QJRAS}}             
\def\skytel{\aaref@jnl{S\&T}}             
\def\ssr{\aaref@jnl{Space~Sci.~Rev.}}     
\def\zap{\aaref@jnl{ZAp}}                 
\def\nat{\aaref@jnl{Nature}}              
\def\aplett{\aaref@jnl{Astrophys.~Lett.}} 
\def\apspr{\aaref@jnl{Astrophys.~Space~Phys.~Res.}} 
\def\physrep{\aaref@jnl{Phys.~Rep.}}      
\def\physscr{\aaref@jnl{Phys.~Scr}}       
\def\commat{\aaref@jnl{Comm.~Math.~Phys.}}              
\def\science{\aaref@jnl{Science}}               
\def\cqg{\aaref@jnl{Classical Quant.~Grav.}}            
\def\jpcs{\aaref@jnl{JPCS}}                                     
\def\ijmpd{\aaref@jnl{Int.~J.~Mod.~Phys.~D}}                    
\def\grg{\aaref@jnl{Gen.~Relat.~Gravit.}}               
\def\rpp{\aaref@jnl{Rep.~Prog.~Phys.}}          
\def\npa{\aaref@jnl{Nucl.~Phys.~A}}        
\def\lrr{\aaref@jnl{Living Rev.~Rel.}}                   
\def\jcap{\aaref@jnl{J.~Cosmology Astropart.~Phys.}}    
\def\rmp{\aaref@jnl{Rev.~Mod.~Phys.}}   

\allowdisplaybreaks[1]

\begin{document}

\title{Rapidly rotating neutron stars in scalar-tensor theories of gravity}

\author{Daniela D. Doneva}
\email{daniela.doneva@uni-tuebingen.de}
\affiliation{Theoretical Astrophysics, Eberhard Karls University of T\"ubingen, T\"ubingen 72076, Germany}
\affiliation{INRNE - Bulgarian Academy of Sciences, 1784  Sofia, Bulgaria}

\author{Stoytcho S. Yazadjiev}
\affiliation{Department of Theoretical Physics, Faculty of Physics, Sofia University, Sofia 1164, Bulgaria}
\affiliation{Theoretical Astrophysics, Eberhard Karls University of T\"ubingen, T\"ubingen 72076, Germany}

\author{Nikolaos Stergioulas}
\affiliation{Department of Physics, Aristotle University of Thessaloniki, Thessaloniki 54124, Greece}

\author{Kostas D. Kokkotas}
\affiliation{Theoretical Astrophysics, Eberhard Karls University of T\"ubingen, T\"ubingen 72076, Germany}
\affiliation{Department of Physics, Aristotle University of Thessaloniki, Thessaloniki 54124, Greece}


\begin{abstract}
We present the field equations governing the equilibrium of rapidly
rotating neutron stars in scalar-tensor theories of gravity, as well
as representative numerical solutions. The conditions for the presence
of a nontrivial scalar field and the deviations from the  general
relativistic solutions are studied. Two examples of scalar-tensor
theories are examined -- one case that is equivalent to the Brans-Dicke
theory and a second case, that is perturbatively equivalent to
Einstein's General Relativity in the weak field regime, but can
differ significantly for strong fields. Our numerical results show
that rapidly rotating neutron star models with a nontrivial scalar
field exist in both cases and that the effect of the scalar
field is stronger for rapid rotation. If we consider values
of the coupling parameters in accordance with current observations,
only the second example of scalar-tensor theories has significant influence
on the neutron star structure. We show that scalarized, rapidly rotating
neutron stars exist for a larger range of the parameters than in the
static case, since  a nontrivial scalar field is present even for
values of the coupling constant $\beta>-4.35$,
and that these solutions are energetically more favorable
than the general relativistic ones. In addition, the deviations
of the rapidly rotating scalar-tensor neutron stars from the
 general-relativistic solutions is significantly larger than
in the static case.
\end{abstract}


\maketitle

\section{Introduction}

Einstein's General Relativity theory (GR) has been tested in various
astrophysical scenarios and its agreement with observations is remarkable.
But, many generalizations of GR exist, which are also compatible with
current observational uncertainties. Moreover, there exist phenomena,
such as dark matter and the accelerated expansion of the universe,
which do not fit well in the standard GR framework and alternative
theories of gravity are often employed to explain them. Alternative
theories of gravity also originate from theories trying to unify all
the interactions such as Kaluza-Klein theories, higher dimensional
theories of gravity, etc. (see \cite{Fujii2003,Capozziello2010,Will2006,Damour1992}
and references therein).

The mediator of the gravitational interaction in GR is the metric
tensor $g_{\mu \nu}$. One of the most natural generalizations of
GR is to include an additional mediator -- a scalar field $\Phi$. In fact, there exists a whole class of \textit{scalar-tensor theories of gravity} (STT), which originate from the works of Jordan, Fierz and Brans \& Dicke \cite{Jordan49,Fierz56,Jordan59,brans611,Dicke62}. In STT, there is no direct interaction between the sources of gravity and the scalar field, which means that the weak equivalence principle is satisfied. To test these alternative theories, it is interesting to explore  solutions representing relativistic stars in STT (in the region of  parameter space that is compatible with current observations) and investigate their deviation from corresponding solutions in GR.

Because of their high densities and compactness, neutron stars are
an ideal laboratory for testing gravity in the strong field regime.
In contrast, most current astrophysical constraints on gravity
theories are derived from observations in weak field environments.
Specifically,  observations of neutron stars may test strong-field
deviations from GR, for STT that are indistinguishable from GR in
the weak field regime \cite{Damour1993,Damour1996}. In the static spherically
symmetric case, it is already known that there exists
\textit{nonuniqueness} of solutions representing neutron stars
for a particular class of STT and a certain range of parameters:
in addition to the general-relativistic solutions with a trivial
scalar field, solutions with a nontrivial scalar field exist,
which are energetically more favorable than their GR counterpart
\cite{Damour1992,Damour1993,Damour1996,Sotani2012,Pani2011a,Salgado1998} and their stability was studied in \cite{Harada1997,Harada1998}.
Under this \textit{spontaneous scalarization}, neutron star
models can have significantly different properties than in GR.
Similarly, black hole solutions in STT with a nontrivial  scalar
field were found and their dynamics was studied
in \cite{Stefanov2008,Doneva2010,Stefanov2007,Georgieva2011,Healy2011,Pani2011,Berti2013,Cardoso2013}.

Astrophysical implications, such as the effect of a nontrivial
scalar field on gravitational wave emission and the redshift
of spectral lines in X-rays and $\gamma$-rays have been
considered in \cite{Sotani04,Sotani2005,DeDeo2003}. Neutron
star mergers in scalar-tensor theories of gravity were studied
in \cite{Barausse2013} and the collapse of a  spherical neutron
star to a black hole was examined in \cite{Novak1998}. The
transition of a neutron star with a zero scalar field to a
scalarized state was considered in  \cite{Novak1998a} and the scalar gravitational waves from Oppenheimer-Snyder collapse
were examined in \cite{Harada1997a}. All
these studies showed that the scalar field has distinct properties,
some of which could potentially be used as observational probes.
However, so far all studies were limited to either static, or
slowly-rotating models  (to first order in the angular velocity)
\cite{Damour1996,Sotani2012}. In the latter case, rotational
corrections to the mass, radius and scalar field, were not
obtained, because they are of higher order with respect to the
rotational frequency. The extension of these studies to rapid
rotation is important from an astrophysical point of view,
because accreting neutron stars in Low-Mass-X-Ray-Binaries
(LMXBs) are known to rotate with frequencies up to 700 Hz
\cite{Hessels2006},
while progenitors of magnetars are theorized to have been born
rapidly rotating \cite{Metzger2011,Giacomazzo2013}. In addition, in binary neutron star
mergers a quasi-stable merger remnant with rapid
differential rotation is now considered as the generic outcome \cite{Faber2012,Stergioulas2011,Read2013}.

Here, we present the first study of rapidly rotating neutron stars in scalar-tensor theories of gravity. The required field equations are derived and they are solved numerically by using a modification of the {\tt rns} code \cite{Stergioulas95}. Our aim is  to check for the existence of solutions with a nontrivial scalar field for two different examples of STT and for different values of the parameters, and to determine the deviations from GR.
For STT, current observations set tight constraints on the possible values of the scalar field coupling constants \cite{Will2006,Damour1996,Damour1998,Freire2012}. We find that the range of parameters for which  scalarized rapidly  rotating neutron star models exist is considerably larger than in the nonrotating case. The strength of the scalar field also increases when rapid rotation is considered  and several stellar properties,  such as mass, radius and angular momentum change significantly.

The paper is structured as follows. In Sec. \ref{Sec:BasicEquation} we give the basic theoretical background and present the field equations. A brief overview of the numerical method, the changes induced by the presence of the scalar field and the various tests we apply in order to check our code are given in Sec. \ref{Sec:NumericalMethod}. In Sec. \ref{Sec:Results} we present the numerical solutions describing rapidly rotating neutron stars with a nontrivial scalar field. Two examples of STT are considered: one, which is equivalent to the Bans-Dicke theory, and a second one, which is perturbatively equivalent to GR in the weak field regime, but can differ significantly for strong fields. We end the paper with a summary
and discussion.

\section{Basic equations} \label{Sec:BasicEquation}

The general form of the gravitational action of the  scalar-tensor
theories in the physical Jordan frame is given by \cite{Fujii2003,Capozziello2010,Will2006,Damour1992}

\begin{eqnarray} \label{JFA}
S = {1\over 16\pi G_{*}} \int d^4x \sqrt{-{\tilde
g}}\left[{F(\Phi)\tilde R} - Z(\Phi){\tilde
g}^{\mu\nu}\partial_{\mu}\Phi
\partial_{\nu}\Phi   -2 U(\Phi) \right] +
S_{m}\left[\Psi_{m};{\tilde g}_{\mu\nu}\right] .
\end{eqnarray}
Here, $G_{*}$ is the bare gravitational constant, and ${\tilde R}$ is
the Ricci scalar curvature with respect to the space-time metric
${\tilde g}_{\mu\nu}$. The dynamics of the scalar field $\Phi$ is
governed by  the functions\footnote{In fact, the three functions
governing the dynamics of the scalar field $\Phi$
 can  be reduced to two functions by a simple redefinition of the scalar field.
 For example, a widely used parametrization is the Brans-Dicke one given by $F(\Phi)=\Phi$ and
 $Z(\Phi)=\frac{\varpi(\Phi)}{\Phi}$. In this parametrization the functions $F(\Phi)$ and $Z(\Phi)$  are reduced to one function, namely $\varpi(\Phi).$}
 $F(\Phi)$, $Z(\Phi)$ and $U(\Phi)$. In
order for the gravitons  to carry positive energy we must have
$F(\Phi)>0$. The nonnegativity of the scalar field kinetic energy
requires that $2F(\Phi)Z(\Phi) + 3[dF(\Phi)/d\Phi]^2 \ge 0$. The
matter fields are collectively denoted by $\Psi_{m}$ and their
action $S_{m}$ depends on $\Psi_{m}$ and the space-time metric
${\tilde g}_{\mu\nu}$. Here we consider the phenomenological case
when the matter action does not involve the scalar field in order
for the weak equivalence principle to be satisfied. \footnote{In
general we can consider the more general case when there are direct
interactions between the matter fields. However we leave this for
future study.}

From a mathematical point of view it is convenient to analyze the scalar-tensor theories with respect to the
conformally related Einstein frame  given by the metric:

\begin{equation}\label {CONF1}
g_{\mu\nu} = F(\Phi){\tilde g}_{\mu\nu} .
\end{equation}

Introducing the scalar field $\varphi$ via the equation

\begin{equation}\label {CONF2}
\left(d\varphi \over d\Phi \right)^2 = {3\over
4}\left({d\ln(F(\Phi))\over d\Phi } \right)^2 + {Z(\Phi)\over 2
F(\Phi)}
\end{equation}
and defining

\begin{equation}\label {CONF3}
{\cal A}(\varphi) = F^{-1/2}(\Phi) \, , \,\, 2V(\varphi) = U(\Phi)F^{-2}(\Phi),
\end{equation}
the Einstein frame version of the  action (\ref{JFA}) takes the form

\begin{eqnarray}
S= {1\over 16\pi G_{*}}\int d^4x \sqrt{-g} \left(R -
2g^{\mu\nu}\partial_{\mu}\varphi \partial_{\nu}\varphi -
4V(\varphi)\right)+ S_{m}[\Psi_{m}; {\cal A}^{2}(\varphi)g_{\mu\nu}],
\end{eqnarray}
where $R$ is the Ricci scalar curvature with respect to the Einstein
metric $g_{\mu\nu}$. By varying this action with respect to the
Einstein frame metric $g_{\mu\nu}$  and the scalar field $\varphi$,
we find the field equations in the Einstein frame
\begin{eqnarray} \label{EFFE}
R_{\mu\nu} - {1\over 2}g_{\mu\nu}R = 8\pi G_{*} T_{\mu\nu}
 + 2\partial_{\mu}\varphi \partial_{\nu}\varphi   -
g_{\mu\nu}g^{\alpha\beta}\partial_{\alpha}\varphi
\partial_{\beta}\varphi -2V(\varphi)g_{\mu\nu}  \,\,\, ,\nonumber
\end{eqnarray}

\begin{eqnarray}
 \nabla^{\mu}\nabla_{\mu}\varphi = - 4\pi G_{*} k(\varphi)T
+ {dV(\varphi)\over d\varphi}
\end{eqnarray}
where
\begin{equation}
k(\varphi)= \frac{d\ln({\cal  A}(\varphi))} {d\varphi}.
\end{equation}

The Einstein frame energy-momentum tensor $T_{\mu\nu}$  is related to
the Jordan frame one ${\tilde T}_{\mu\nu}$ via $T_{\mu\nu}= {\cal
A}^2(\varphi){\tilde T}_{\mu\nu}$. In the case of a perfect fluid,
the Jordan frame energy-momentum tensor is

\begin{eqnarray}
{\tilde T}_{\mu\nu}= (\tilde \varepsilon + \tilde p){\tilde u}_{\mu}
{\tilde u}_{\nu} + {\tilde p} {\tilde g}_{\mu\nu}
\end{eqnarray}
while its form in the Einstein frame is given by

\begin{eqnarray}
T_{\mu\nu}= (\varepsilon + p)u_{\mu} u_{\nu} + pg_{\mu\nu}.
\end{eqnarray}
The relations between the quantities in both frames are explicitly
given by the following equations:

\begin{eqnarray}\label{DPTEJF}
\varepsilon &=&{\cal A}^4(\varphi){\tilde\varepsilon}, \nonumber \\
p&=&{\cal A}^4(\varphi){\tilde p},  \\
u_{\mu}&=& {\cal A}^{-1}(\varphi){\tilde u}_{\mu} . \nonumber
\end{eqnarray}

The contracted Bianchi identities give the following conservation
law for the Einstein frame energy-momentum tensor

\begin{eqnarray}
\nabla_{\mu}T^{\mu}{}_{\nu} = k(\varphi)T\partial_{\nu}\varphi \,\,\,
. \nonumber
 \end{eqnarray}

In accordance with the main purpose of the present paper we consider
stationary and  axisymmetric spacetimes. In mathematical terms this
means that the spacetimes admit one asymptotically timelike at infinity Killing
vector field $\xi$ and another  axial spacelike Killing vector field
$\eta$ with periodic orbits and commuting with $\xi$ such that their flows leave invariant
not only the metric (i.e. $\pounds_{\xi}
g_{\mu\nu}=\pounds_{\eta}g_{\mu\nu}$=0) but also the scalar field
and the matter fields, i.e.

\begin{eqnarray} \label{LieDT}
\pounds_{\xi}\varphi= \pounds_{\eta}\varphi=0, \\
 \pounds_{\xi}T_{\mu\nu}=
\pounds_{\eta}T_{\mu\nu}=0,
\end{eqnarray}
with $\pounds$ being the Lie derivative. It is rather natural from a
physical point of view to require   the set of fixed points under
the flow of $\eta$ to be nonempty. In other words we require $\eta$
to have a nontrivial axis of symmetry where $\eta$ vanishes.

We should note that in the Jordan frame we also have
$\pounds_{\xi}{\tilde g}_{\mu\nu}=\pounds_{\eta}{\tilde
g}_{\mu\nu}=0$ and $\pounds_{\xi}{\tilde
T}_{\mu\nu}=\pounds_{\eta}{\tilde T}_{\mu\nu}=0$ since the scalar
field $\varphi$ is invariant under the Killing flows.

We will impose one more condition, namely the stationary and
axisymmetric spacetime to be circular. In other words, the
2-dimensional surfaces orthogonal to the Killing fields $\xi$ and
$\eta$ to be integrable. According to the Frobenius theorem this
holds if and only if the Frobenius conditions are satisfied, namely
\cite{Kundt1966,Carter1969}

\begin{eqnarray}
\eta_{[\mu}\xi_{\nu}\nabla_{\alpha}\xi_{\beta]}=0 , \\
\xi_{[\mu}\eta_{\nu}\nabla_{\alpha}\eta_{\beta]}=0.
\end{eqnarray}

Using the properties of the Killing fields and the field equations,
the Frobenius conditions can be written in the form

\begin{eqnarray}
\xi^{\mu} T_{\mu[\nu}\xi_{\alpha}\eta_{\beta]}=0, \\
\eta^{\mu}T_{\mu[\nu}\eta_{\alpha}\xi_{\beta]}=0.
\end{eqnarray}
Written in this form, the Frobenius conditions themselves  impose
restrictions on the energy-momentum tensor. Specifically, for the
case of a perfect fluid  we obtain (taking into account that
$\xi^{\mu}u_{\nu}\ne 0$ and $\varepsilon + p\ne 0$)
\begin{eqnarray}
u_{[\nu}\xi_{\alpha}\eta_{\beta]}=0,
\end{eqnarray}
which means that at every spacetime point the 4-velocity $u^{\mu}$
lies in the plane spanned by $\xi$ and $\eta$, i.e.
$u^{\mu}=u^{\xi}\xi^{\mu} + u^{\eta}\eta^{\mu}$. This also holds in
the physical Jordan frame, i.e. ${\tilde u}^{\mu}={\tilde
u}^{\xi}\xi^{\mu} + {\tilde u}^{\eta}\eta^{\mu}$ since
$\pounds_{\xi,\eta}\varphi=0$.

Using the fact that $u$ is a linear combination of $\xi$ and
$\eta$ as well as  that $\xi$ and $\eta$ commute, we obtain from
(\ref{LieDT}) the natural consequence

\begin{eqnarray}
\pounds_{\xi,\eta} \varepsilon= \pounds_{\xi,{\eta}} p=
\pounds_{\xi,\eta} {\tilde \varepsilon}= \pounds_{\xi,\eta} {\tilde
p}=0.
\end{eqnarray}

The circularity condition allows us to simplify considerably the
spacetime metric. The circular spacetimes admit a foliation by
integrable 2-surfaces orthogonal to the Killing fields $\xi$ and
$\eta$ and the spacetime is (locally) a product manifold
$M=M_{\|}\times M_{\perp}$. Here $M_{\|}$ is a 2-dimensional Lorentz
manifold to which the Killing fields $\xi$ and $\eta$ are tangential
and $M_{\perp}$ is a 2-dimensional Riemannian manifold orthogonal to
$\xi$ and $\eta$. The spacetime metric is then given by the sum
$g=g_{\|} + g_{\perp}$ where $g_{\|}$ is a Lorentz metric on
$M_{\|}$ and $g_{\perp}$ is a Riemannian metric on $M_{\perp}$. In a
coordinate presentation we have

\begin{eqnarray}
ds^2= g_{\mu\nu}dx^\mu dx^\nu= g_{\|ab}dx^adx^b + g_{\perp\, ij}
dx^idx^{j},
\end{eqnarray}
where both metrics depend only on the coordinates $x^{k}$ on
$M_{\perp}$, namely $g_{\|ab}=g_{\|ab}(x^k)\;$, $\;\;g_{\perp\, ij}=g_{\perp\,
ij}(x^k) $.

Now we can choose convenient coordinates on $M_{\perp}$ and $M_{\|}$
which automatically fix convenient spacetime coordinates.  Since
$(M_{\perp}, g_{\perp\, ij})$ is a two-dimensional Riemannian
manifold we can always chose the coordinates $(r, \theta)$ so that
the metric has the conformal from

\begin{eqnarray}
 g_{\perp\, ij}dx^i dx^j=e^{2\alpha} (dr^2 + r^2d\theta^2)
\end{eqnarray}
with $\alpha=\alpha(r, \theta)$. Concerning the 2-dimensional
Lorentz manifold $(M_{\|}, g_{\|ab})$, the natural choice of the
coordinates on it are the coordinates adapted to the Killing fields,
i.e. the coordinates $t$ and $\phi$ for which
$\xi=\frac{\partial}{\partial t}$ and $\eta=
\frac{\partial}{\partial \phi}$. Denoting by $\omega$ the minus
scalar product of the Killing fields normalized by the norm of
$\eta$, i.e. $\omega= -g(\xi, \eta)/g(\eta,\eta)$, the Lorentz metric
can be written in the form

\begin{eqnarray}
g_{\|ab}dx^a dx^b= - A^2 dt^2 + B^2r^2\sin^2\theta(d\phi - \omega
dt)^2
\end{eqnarray}
with $\omega=\omega(r,\theta)$, $A=A(r,\theta)$ and $B(r, \theta)$.
It turns out that it is more convenient in dealing with the field
equations to use the functions $\sigma$ and $\gamma$ instead of $A$ and
$B$, defined by $A^2=e^{\sigma + \gamma}$ and $B^2= e^{\gamma -
\sigma}$. Summarizing, we have chosen coordinate $t,\phi, r, \theta $ in
which the spacetime metric takes the form

\begin{eqnarray}
&&ds^2 = -e^{\gamma+\sigma} dt^2 + e^{\gamma-\sigma} r^2
\sin^2\theta (d\phi - \omega dt)^2 + e^{2\alpha}(dr^2 + r^2
d\theta^2)
\end{eqnarray}
with all metric functions depending only on $r$ and $\theta$.

Having the explicit form of the metric, we proceed to write the
dimensionally reduced field equations. For this purpose it is
convenient to use the proper velocity $v$ of the fluid given by

\begin{equation}
v = (\Omega - \omega) r \sin \theta e^{-\sigma},
\end{equation}
where $\Omega$ is the fluid angular velocity defined by
$\Omega=\frac{u^{\phi}}{u^{t}}$. $\Omega$ and $v$ are the same in
both Einstein and Jordan frames. Indeed, for the angular velocity
${\tilde \Omega}$ in the Jordan frame we have ${\tilde
\Omega}=\frac{{\tilde u}^{\phi}}{{\tilde u}^{t}}=\frac{{\cal
A}(\varphi) u^{\phi}}{{\cal A}(\varphi)
u^{t}}=\frac{u^{\phi}}{u^{t}}=\Omega$. In the same way the factor
${\cal A}(\varphi)$ cancels out in the definition of the proper
velocity ${\tilde v}$ in the Jordan frame which gives that ${\tilde
v}=v$.

 The fluid four
velocity in the Einstein frame then is
\begin{equation}
u^\mu = \frac{e^{-(\sigma + \gamma)/2}}{\sqrt{1-v^2}}
[1,0,0,\Omega].
\end{equation}

From a numerical point of view it is more convenient to use the
following angular coordinate $\mu=\cos\theta$ instead  of $\theta$.
With the help of this coordinate, the dimensionally reduced Einstein
equations for the metric functions $\gamma$, $\sigma$ and $\omega$
are the following:

\begin{eqnarray}
\left(\Delta + \frac{1}{r} \partial_{r}  -
\frac{\mu}{r^2}\partial_{\mu}\right)\left(\gamma
e^{\gamma/2}\right)&=&e^{\gamma/2}\left\{(16\pi p -
4V(\varphi))e^{2\alpha}  + \right. \notag \\
&& \notag \\
&&\left.+\frac{\gamma}{2}\left[(16\pi p -4V(\varphi))e^{2\alpha} -
\frac{1}{2}(\partial_{r}\gamma)^2 - \frac{1}{2} \frac{1-\mu^2}{r^2}
(\partial_{\mu}\gamma)^2 \right]\right\}, \label{eq:DiffEq_gamma}
\end{eqnarray}

\begin{eqnarray}
\Delta(\sigma e^{\gamma/2}) &=& e^{\gamma/2}\left\{8\pi (\varepsilon
+ p)e^{2\alpha} \frac{1+ \upsilon^2}{1 - \upsilon^2}   + r^2
(1-\mu^2)e^{-2\sigma}\left[(\partial_{r}\omega)^2  +
\frac{1-\mu^2}{r^2}(\partial_{\mu}\omega)^2\right] +
\frac{1}{r}\partial_{r}\gamma - \frac{\mu}{r^2}\partial_{\mu}\gamma
 \right. \nonumber  \\ && \nonumber \\
 &&\left.  + \frac{\sigma}{2}\left[(16\pi p -4V(\varphi))e^{2\alpha} - \frac{1}{r}\partial_{r}\gamma + \frac{\mu}{r^2}\partial_{\mu}\gamma -
 \frac{1}{2}(\partial_{r}\gamma)^2  - \frac{1}{2}\frac{1-\mu^2}{r^2} (\partial_{\mu}\gamma)^2 \right] \right\}, \label{eq:DiffEq_sigma}
\end{eqnarray}

\begin{eqnarray}
\left(\Delta  +  \frac{2}{r} \partial_{r}  -
\frac{2\mu}{r^2}\partial_{\mu}\right)\left(\omega e^{\gamma/2
-\sigma}\right) &=& e^{\gamma/2 - \sigma} \left\{- 16\pi
\frac{(\varepsilon + p)(\Omega - \omega)}{1- \upsilon^2}e^{2\alpha}
+ \right. \nonumber \\
&&+\omega\left[-\frac{1}{r}\partial_{r} (\frac{1}{2}\gamma + 2\sigma)
+ \frac{\mu}{r^2}\partial_{\mu}(\frac{1}{2}\gamma + 2\sigma) -
\frac{1}{4}(\partial_{r}\gamma)^2 -  \frac{1}{4}\frac{1-\mu^2}{r^2}
(\partial_{\mu}\gamma)^2  + \right. \nonumber \\
&& + (\partial_{r}\sigma)^2 + \frac{1-\mu^2}{r^2}
(\partial_{\mu}\sigma)^2     - r^2(1-\mu^2)e^{-2\sigma}
\left((\partial_{r}\omega)^2 + \frac{1-\mu^2}{r^2}
(\partial_{\mu}\omega)^2\right)  \nonumber \\
&& \left.  \left. - 8\pi \frac{\varepsilon (1+ \upsilon^2) +
2p\upsilon^2}{1- \upsilon^2} e^{2\alpha} -2 V(\varphi)
e^{2\alpha}\right] \right\}. \label{eq:DiffEq_omega}
\end{eqnarray}

Here, the differential operator $\Delta$ is defined by
\begin{eqnarray}
\Delta =  \partial^2_{r} + \frac{2}{r}\partial_{r} +
\frac{1-\mu^2}{r^2}\partial^2_{\mu} - \frac{2\mu}{r^2}
\partial_{\mu}.
\end{eqnarray}

For the metric function $\alpha$ we obtain two first order partial
differential equations. However, in the numerical method in order to
determine $\alpha$ we need only one of them, namely the following

\begin{eqnarray}
&&\partial_\mu \alpha = -\frac{\partial_\mu \gamma + \partial_\mu \sigma}{2} - \left\{(1-\mu^2)(1+r\partial_r\gamma)^2 + [-\mu + (1-\mu^2)\partial_\mu \gamma]^2 \right\}^{-1} \times  \label{eq:DiffEq_alpha}\\ \notag \\
&&\left\{\frac{1}{2}\left[r\partial_r(r\partial_r\gamma) + r^2 (\partial_r \gamma)^2 - (1-\mu^2)(\partial_\mu \gamma)^2 - \partial_\mu[(1-\mu^2)\partial_\mu \gamma] + \mu\partial_\mu \gamma\right] \times [-\mu + (1-\mu^2)\partial_\mu \gamma] + \right. \notag \\ \notag \\
&& +\frac{1}{4}[-\mu + (1-\mu^2)\partial_\mu \gamma] \times \left[r^2(\partial_r \gamma + \partial_r \sigma)^2 - (1-\mu^2)(\partial_\mu \gamma + \partial_\mu \sigma)^2 + 4r^2(\partial \varphi)^2 - 4(1-\mu^2)(\partial_\mu \varphi)^2\right] + \notag \\ \notag \\
&& + \mu r \partial_r\gamma [1+r\partial_r\gamma] - (1-\mu^2)r(1+r\partial_r\gamma)\left[\partial_\mu\partial_r \gamma + \partial_\mu\gamma\partial_r\gamma + \frac{1}{2}(\partial_\mu \gamma + \partial_\mu \sigma)(\partial_r \gamma + \partial_r \sigma) + 2 \partial_\mu\varphi\partial_r \varphi \right] +\notag \\ \notag \\
&&+\frac{1}{4}(1-\mu^2)e^{-2\sigma}\left[-[-\mu+(1-\mu^2)\partial_\mu
\gamma][r^4(\partial_r \omega)^2 - r^2(1-\mu^2)(\partial_\mu
\omega)^2] + \right. \notag \\ \notag \\
&&\left.\left.+2(1-\mu^2)r^3\partial_\mu\omega\partial_r\omega(1+r\partial_r\gamma)\right]\right\}.
\notag
\end{eqnarray}

The above equations have to be supplemented with the field equation
for the scalar field and the equation for hydrostatic equilibrium
\begin{eqnarray}
\Delta \varphi= - \partial_{r}\gamma\partial_{r}\varphi -
\frac{1-\mu^2}{r^2} \partial_{\mu}\gamma\partial_{\mu}\varphi +
\left[4\pi k(\varphi)(\varepsilon - 3p) +
\frac{dV(\varphi)}{d\varphi}\right]e^{2\alpha},
\label{eq:DiffEq_phi}
\end{eqnarray}

\begin{eqnarray}
\frac{\partial_i{\tilde p}}{{\tilde \varepsilon} + \tilde{p}} -
\left[\partial_i(\ln \, u^t) - u^t u_\phi \partial_i \Omega -
k(\varphi) \partial_i \varphi\right]=0 \label{eq:Hydrostatic_Equil}.
\end{eqnarray}

Here we use the Jordan (physical) frame pressure $\tilde{p}$ and
energy density ${\tilde \varepsilon}$ which are related to the
Einstein frame ones via (\ref{DPTEJF}). On the other hand, the Einstein frame four velocity is utilized because the equations take a simpler form in this case.

To close the system describing the equilibrium
configurations of rapidly rotating neutron stars in the
scalar-tensor theories we should specify the equations of state for
the matter. In the present paper we use a polytropic
equation of state (EoS):
\begin{eqnarray}
&& {\tilde \varepsilon} = K \frac{{\tilde \rho}^\Gamma}{\Gamma-1} + {\tilde \rho} c^2, \notag \\
&& {\tilde p} = K {\tilde \rho,}^\Gamma \label{eq:EOS} \\
&& \Gamma = 1+ \frac{1}{N}, \notag
\end{eqnarray}
where ${\tilde \rho}$ is the rest mass density in the Jordan frame, $N$
is the polytropic index and $K$ is the polytropic constant.

The global physical characteristics of the rapidly rotating neutron
stars in which we are interested, are the mass $M$, the angular
momentum $J$, the total baryon rest mass $M_0$ and the
circumferential radius of the star $R_e$ in the physical Jordan frame.
In contrast  with general relativity, the definition of mass in
scalar-tensor theories of gravity is in general quite subtle because
scalar-tensor theories  violate in general the strong equivalence
principle. This results in the appearance of different possible
masses as a measure of the total energy of the star
\cite{Lee1974,Scheel1995, Scheel1995a, Whinnett1999,Yazadjiev1999}.
It turns out that only the so-called ``tensor'' mass has natural
energy-like properties; for example the tensor mass is positive
definite, it decreases monotonically by the emission of gravitational waves
and it is well defined even for dynamical spacetimes
\cite{Lee1974,Scheel1995,Scheel1995a}. Moreover, it is the tensor
mass that leads to a physically acceptable picture in the theory of
the stars -- the tensor mass peaks at the same point as the particle
number, which is a crucial property for the stability of the static
stars \cite{Whinnett1999,Yazadjiev1999}. Therefore, it is the tensor
mass that should be taken as the physical mass\footnote{In some
specific scalar-tensor theories it is possible for  the ADM masses
defined in both frames to coincide. This is the case for example
with the second  scalar-tensor theory considered in the present
paper. Because of the special dependence  ${\cal
A}(\varphi)=\exp(1/2\beta\varphi^2)$ and the condition
$\lim_{r\to\infty}\varphi=0$, it is not difficult to show that
both frame metrics have the same leading asymptotic which results in
the coincidence of the masses. }. By definition the tensor mass is
just the Arnowitt-Deser-Misner (ADM) mass in the Einstein frame. As a direct consequence of
this definition and using the Komar integral we find the following
expression for the tensor mass:
\begin{eqnarray}
M=\int_{Star}\left[(\varepsilon + 3p)  + 2(\varepsilon +
p)\frac{\Omega}{\Omega-\omega} \frac{\upsilon^2}{1-\upsilon^2} -
\frac{1}{2\pi}V(\varphi)\right] \sqrt{-g}d^3x. \label{eq:TotalMass}
\end{eqnarray}

The angular momentum $J$ is defined as usual and it is the same in
both frames. The Komar integral  gives the following equation for the
angular momentum expressed in terms of the Einstein frame quantities:

\begin{eqnarray}
J= \int_{Star}(\varepsilon + p) \frac{\upsilon^2}{1-\upsilon^2}
\frac{1}{\Omega- \omega} \sqrt{-g} d^3x. \label{eq:AngularMomentum}
\end{eqnarray}

Concerning the total baryon rest mass, the equation in the physical  Jordan
frame is the usual one

\begin{eqnarray}
M_0=\int_{Star} {\tilde \rho} {\tilde u}^{\mu}
d\Sigma_{\mu}=\int_{Star} {\tilde \rho} {\tilde u}^{t}\sqrt{-{\tilde
g}} d^3x ,
\end{eqnarray}
where ${\tilde \rho}$  is the rest mass density in the Jordan frame. The above equation, expressed in terms of the
four-velocity and metric in the
Einstein frame, takes the form
\begin{eqnarray}
M_0=\int_{Star} {\cal A}^{3}(\varphi) {\tilde \rho} u^{t}
\sqrt{-g} d^3x.
\end{eqnarray}

The circumferential radius ${\tilde R}_{e}$ of the star in the
physical Jordan frame is defined to be  the normalized by $2\pi$
circumference of the circle in the equatorial plane where the
pressure vanishes, i.e.
\begin{equation}
{\tilde R}_e = {\cal A}(\varphi)\; r \;
e^{(\gamma-\sigma)/2}|_{r=r_{e},\theta=\pi/2},
\end{equation}
where $r_e$ is the coordinate equatorial radius of the star given by
${\tilde p}(r_{e},\theta=\pi/2)=0$.

The Kepler angular velocity $\Omega_K$ of a neutron star is defined as the angular velocity of a free particle in circular orbit in the $\theta=\pi/2$ plane. The mass-shedding or Kepler limit along a sequence of rotating stellar models is reached when the angular velocity of the neutron star fluid at the equator is equal to the Kepler angular velocity. In terms of Einstein frame quantities it is given by
\begin{eqnarray}
\Omega_K &=& \left(\omega  + \frac{\omega'}{8+\gamma'-\sigma' + 2 k(\varphi) \varphi'} + \right. \notag \\ \notag \\
&& \left.\left.+ \sqrt{\left[\frac{\omega' }{8 + \gamma' - \sigma' + 2 k(\varphi)\varphi'}  \right]^2  + \frac{e^{2\sigma}(\sigma' + \gamma' + 2 k(\varphi) \varphi')}{r_e^2(8 + \gamma' - \sigma' + 2 k(\varphi) \varphi')}}\right)\right|_{r=r_e,\theta=\pi/2} \label{eq:KeplerOmega}
\end{eqnarray}
where $(')$ stands for $\partial_r$. But as we pointed out, $\Omega$ has the same values in both Einstein and Jordan frames so this
equation gives us also the physical or Jordan frame Kepler angular velocity.

The procedure of introducing dimensionless variables is the same as in \cite{Cook1992}. We set $c=G=1$ and as $K^{N/2}$ has units of
length it can be used as a fundamental length scale of the system.

\section{Numerical method} \label{Sec:NumericalMethod}
We follow the numerical method of H. Komatsu, Y. Eriguchi, and I. Hachisu \cite{Komatsu1989,Komatsu1989a} with the modifications
introduced in \cite{Cook1992}.
As a base we use the {\tt rns} code \cite{Stergioulas95} where the additional terms and equations coming from the scalar field are implemented.
Below we will describe very briefly some of the main points of the method.

First we have to note that for convenience, instead of the radial coordinate $r$ we use a compactified one given by
\begin{equation}
r \equiv r_e \left(\frac{s}{1-s}\right),
\end{equation}
where the integration domain $r \in [0,\infty)$ is mapped to $s \in [0,1)$ and the equatorial radius of the star is always at $s=0.5$.

The essence of the method is that the differential equations \eqref{eq:DiffEq_gamma}--\eqref{eq:DiffEq_omega} are transformed into an integral form using the Green functions which enable us to handle the boundary conditions in a simple manner. The boundary conditions come from the requirements for regularity of the functions at the center of the star and the rotational axes, and asymptotic flatness at infinity. For each differential equation a specific Green function is chosen that fulfils the corresponding boundary conditions as it is explained in detail in \cite{Komatsu1989}. The introduction of a scalar field alters only the source terms in the integral representation of the metric potential equations \eqref{eq:DiffEq_gamma}--\eqref{eq:DiffEq_omega}. But a new second order differential equation for the scalar field appears, that is eq. \eqref{eq:DiffEq_phi}, and we have to derive its integral representation. One can easily conclude that the appropriate Green function in this case is the same as for the metric potential $\sigma$ and after taking into account the equatorial and axial symmetry we obtain
\begin{eqnarray}
\varphi &=& -\sum^\infty_{n=0} P_{2n}(\mu)\left[\left(\frac{1-s}{s}\right)^{2n+1}\int_0^s \frac{ds' s'^{\;2n}}{(1-s')^{2n+2}} \int_0^1 d\mu' P_{2n}(\mu'){\tilde S}_\varphi(s',\mu') +\right. \notag \\ && \notag\\
&&\left. + \left(\frac{s}{1-s}\right)^{2n}\int_s^1 \frac{ds'(1-s')^{2n-1}}{s'^{\;2n+1}} \int_0^1 d\mu' P_{2n}(\mu') {\tilde S}_\varphi(s',\mu') \right],
\end{eqnarray}
where $P_n(\mu)$ are the associated Legendre polynomials. The effective source ${\tilde S}_\varphi(s,\mu)$ is defined as
\begin{eqnarray}
 {\tilde S}_\varphi(s,\mu) &=& - s^2 (1-s)^2 \partial_{s}\gamma\partial_{s}\varphi -
(1-\mu^2) \partial_{\mu}\gamma\partial_{\mu}\varphi + \\ \notag \\
&&+\frac{r_e^2 s^2}{(1-s)^2}e^{2\alpha}\left[4\pi
k(\varphi)(\varepsilon - 3p)
+\frac{dV(\varphi)}{d\varphi}\right].\notag
\end{eqnarray}
The asymptotic behavior at infinity ($\varphi \sim {\cal O}(1/r)$ for large $r$) is secured by the choice of the Green function.

The only differential equation that is not transformed into integral form is the first order differential equation for the metric function $\alpha$ -- Equation \eqref{eq:DiffEq_alpha}. The flatness condition at the rotation axis requires that
\begin{equation}
\alpha = \frac{\gamma-\sigma}{2} \;\;\; {\rm at} \;\;\; \mu=\pm 1,
\end{equation}
and we can integrate equation \eqref{eq:DiffEq_alpha} with the above condition.
The asymptotic flatness of $\alpha$ at infinity is fulfilled automatically, because all other metric potentials,
as well as the scalar field, satisfy the correct boundary conditions at infinity.

The last equation we have to address in detail is the equation for hydrostationary equilibrium given by \eqref{eq:Hydrostatic_Equil}.
Taking into account the standard integrability condition related to the rotational law \cite{Bardeen1970,Butterworth1976}
\begin{equation}
u^t u_\phi = F[\Omega],
\end{equation}
we can transform equation \eqref{eq:Hydrostatic_Equil}  into:
\begin{equation}
H - \ln u^t + \ln {\cal A}(\varphi) + \int_{\Omega_C}^\Omega F[\Omega] d\Omega = {\rm const}, \label{eq:Entropi_Equation}
\end{equation}
where $\Omega_C$ is the angular velocity at the center and $H$ is the specific enthalpy which is defined up to an additive constant as
\begin{equation}
H = \int_0^{\tilde p} \frac{d{\tilde p}}{{\tilde \varepsilon} + {\tilde p}}.
\end{equation}
In the present paper we assume uniform rotation and that is why we can set $F[\Omega]=0$. The differentially rotating case will be considered elsewhere.

The numerical procedure of solving the differential equations is explained in detail in \cite{Komatsu1989,Cook1992,Friedman2013}. An important point is that the ratio of the polar to the equatorial radius $r_p/r_e$ is used as an input parameter instead of the angular velocity of the star. This choice is more suitable, for example, when nonuniqueness of the solutions is present in the case of rapid rotation, or in the case when the matter has toroidal topology instead of spheroidal ($r_e$ is the outer radius of the torus then). The second input parameter is the energy density at the center of the star. The procedure for finding solutions is the standard one -- we have to supply first an initial approximation for the metric functions, the energy density and the scalar field, and after substituting them in equations \eqref{eq:DiffEq_gamma}--\eqref{eq:DiffEq_omega}, \eqref{eq:DiffEq_alpha}--\eqref{eq:Hydrostatic_Equil}, we obtain the new updated values. A solution is obtained when the difference between the values of selected properties (such as the radius of the star) at two consecutive iterations is small enough.

After implementing the required changes, we checked the modified {\tt rns} code successfully against several limiting cases:
\begin{itemize}
\item \textbf{The GR limit.} When we set the potential $V(\varphi)$ and the coupling function $k(\varphi)$ to zero, the code converges toward the general relativistic solutions with trivial scalar field.
\item \textbf{The nonrotating limit.} To be able to make a more profound verification of our code, we created a new one dimensional code that solves the ordinary differential equations describing nonrotating neutron stars in STT. For this purpose it is most convenient to use the equations given in \cite{Damour1992,Damour1993,Damour1996}. Those  equations are much simpler and it is easier to converge to a solution with nontrivial scalar field as ordinary differential equations are solved with only one shooting parameter -- the central value of the scalar field. The comparison between the two codes showed an excellent agreement. Our results also agree well with the results for static neutron stars in STT presented in \cite{Damour1993,Sotani04}.
\item \textbf{The slow rotation approximation (of order $\Omega$).} The rotating scalar-tensor neutron stars in slow rotation approximation were studied in \cite{Damour1996,Sotani2012}. In this approximation the rotational corrections to the mass, radius, scalar field, etc. cannot be calculated, as they are of higher order with respect to the angular velocity of the star, but we can compute for example the $g_{t\phi}$ component of the metric and the angular momentum. To make a systematic comparison, we chose to implement the slow rotation modifications given in \cite{Damour1996} in our one dimensional code. The results produced by {\tt rns} are in a very good agreement with those produced with the slow rotation code.
\item \textbf{The mass and the angular momentum.} There are two independent ways to determine the mass and the angular momentum of the star that are valid also in the rapidly rotating case. First, one can use the
integral Equations \eqref{eq:TotalMass} and \eqref{eq:AngularMomentum}. Alternatively, $M$ and $J$ can be determined from the asymptotic form of the metric components at infinity $g_{tt}|_{r\rightarrow\infty} \approx -(1-2M/r)$ and $g_{t\phi}|_{r\rightarrow\infty} \approx -2 J \sin^2 \theta /r$. The two values of $M$ and $J$ coincide within roughly $0.3\%$ even in the rapidly rotating case.
\end{itemize}

\section{Results}\label{Sec:Results}

In our studies we set the scalar field potential $V(\varphi)$ to
zero. We will consider two representative scalar-tensor theories
given by the following functions\footnote{$k_0$ is usually denoted
by $\alpha$ in most previous publications, but we use a different
notation, as in our case $\alpha$ is reserved for one of the metric
potentials.} ${\cal A}(\varphi)$:
\begin{itemize}
\item $\ln {\cal A}(\varphi) =  k_0 \varphi $,
\item $\ln {\cal A}(\varphi) = \frac{1}{2} \beta \varphi^2$,
\end{itemize}
with zero background value for the scalar field, i.e.
\begin{eqnarray}
\lim_{r\to \infty}\varphi = 0.
\end{eqnarray}

The first case is equivalent to the Brans-Dicke scalar-tensor theory.
In the second case, the STT is perturbatively equivalent to GR
in the weak field regime. For strong fields, though, interesting
new effects can appear (such as a bifurcation due to nonuniqueness of solutions)
that were already observed for static neutron
stars \cite{Damour1993,Damour1996} and
black holes \cite{Stefanov2008,Doneva2010}.
Moreover, typically it is energetically
more favorable for the compact object to possess a nontrivial scalar field. Also, the field equations are invariant under the change of sign of the scalar field, $\varphi \rightarrow -\varphi$, for the second
class of STT. Therefore, two nontrivial branches of solutions exist, which differ only in the sign of the scalar field,
but otherwise have identical global properties, such as mass, radius, angular momentum, etc.

For large values of the parameters $k_0$ and $\beta$, the STT neutron stars can
differ significantly from the general relativistic solutions, but the
experiments set constraints on these parameters,  which currently are
roughly $k_0<4\times 10^{-3}$ and $\beta>-4.8$ \cite{Will2006,Freire2012}.
 On their own, such small values of $k_0$ would lead to a very weak scalar field,
with a negligible effect on the neutron star structure, but the
constraints on $\beta$ allow for the development of a
nontrivial scalar field
(spontaneous scalarization \cite{Damour1993})
which can alter the neutron star properties  significantly.

We will consider a polytropic equation of state given by
Equation \eqref{eq:EOS} with $N=0.7463$ and $K=1186$, which is chosen to
match equation of state II in \cite{DiazAlonso1985} ($K$ is given in
dimensionless units $c=G=M_\odot=1$). It is widely used in the literature on
static neutron stars in STT \cite{Damour1993,Harada1998,Sotani04} and it is
a convenient model for comparisons.

Let us also briefly discuss how we converge to a nontrivial solution: In
the case when $\ln {\cal A}(\varphi) =  k_0 \varphi $,  a unique solution of the field
equations exists for a given central energy density and axis ratio,
and the numerical scheme converges to it, starting from a static GR
solution (i.e. with a trivial scalar field) of the same central energy density. But, in the case with
$\ln {\cal A}(\varphi) = \frac{1}{2} \beta \varphi^2$, three different equilibrium solutions
can exist for a certain range of parameters: one with a trivial scalar field (equivalent to the GR solution)
and two nontrivial solutions with positive and negative signs of the scalar field, but with identical global characteristics given by \eqref{eq:TotalMass}--\eqref{eq:KeplerOmega}. To converge to the nontrivial solutions,
it suffices to use a static GR solution with a nontrivial scalar field as an initial guess
(this solution is found with a new TOV solver in {\tt rns}, solving the equations
given in \cite{Damour1993,Damour1996}). In cases where for a given central
energy density no nontrivial solutions exist in the static limit, but a nontrivial
scalar field is expected in the rapidly rotating case, we use the trivial (GR) static solution
together with a (chosen) constant scalar field as an initial guess. With careful choices for
the initial scalar field, the numerical method in {\tt rns} then also converges to the
nontrivial solutions, when they exist.

\subsection{Scalar-tensor theories with $\ln {\cal A}(\varphi) =  k_0 \varphi$}
This example of STT is actually equivalent to the Brans-Dicke theory.
Here, only one neutron star solution of the field equations exist for a given central energy density and axis ratio
and neutron star models have a
nontrivial scalar field for any nonzero value of $k_0$.
Because of the observational constraint
$k_0<4\times 10^{-3}$, we will consider this case very briefly.

Two sequences of solutions were calculated for the case
$k_0=4\times 10^{-3}$ -- the static one and
a sequence of models rotating at the mass-shedding limit with
angular velocity \eqref{eq:KeplerOmega}.
We find that even though a nontrivial scalar field is present and its value increases for rapid rotation,
it is very  weak and the total mass, radius and angular momentum of
the stars are the same as the corresponding quantities in the
GR limit within the numerical error.

\begin{figure}[ht!]
\centering
\includegraphics[width=0.48\textwidth]{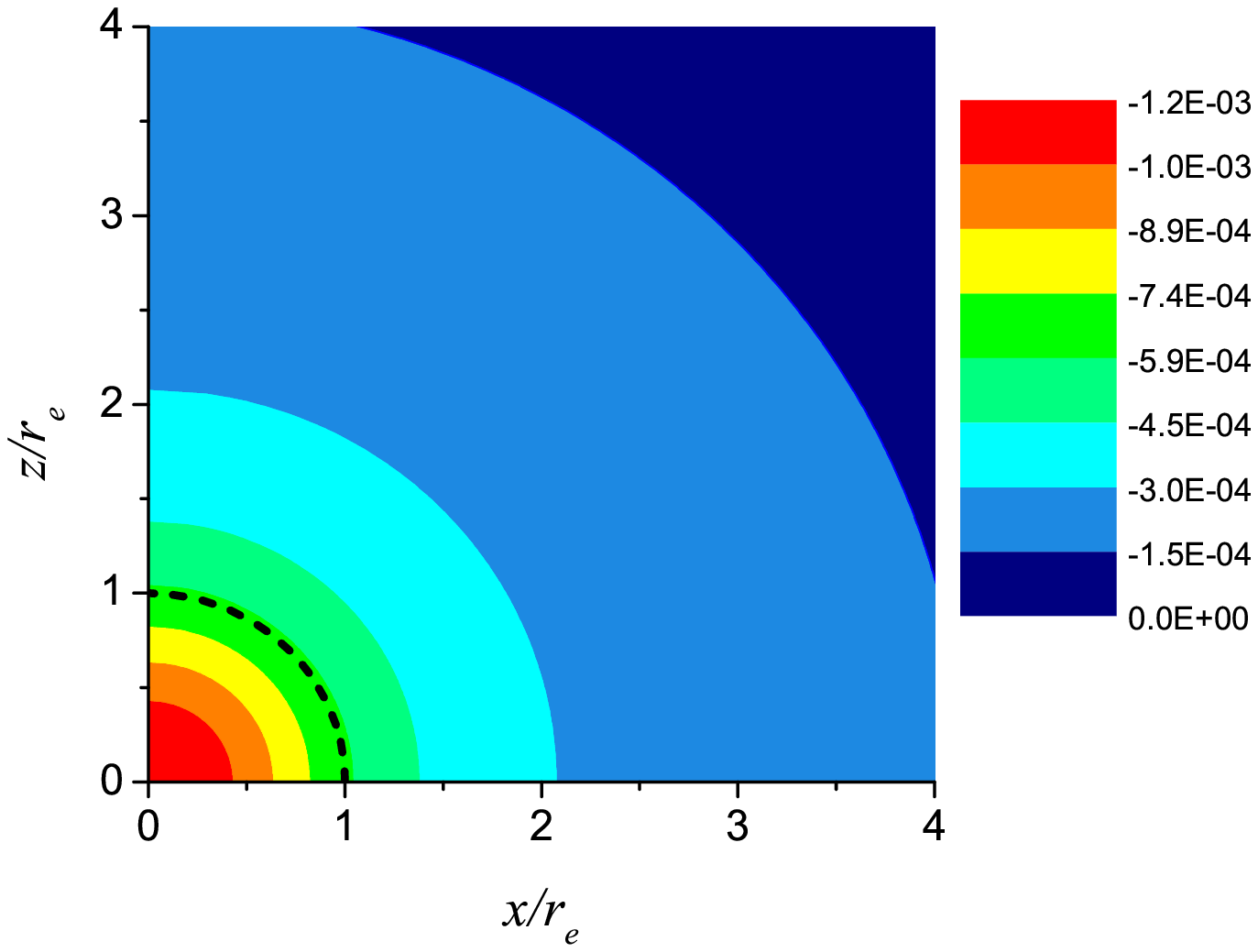}
\includegraphics[width=0.48\textwidth]{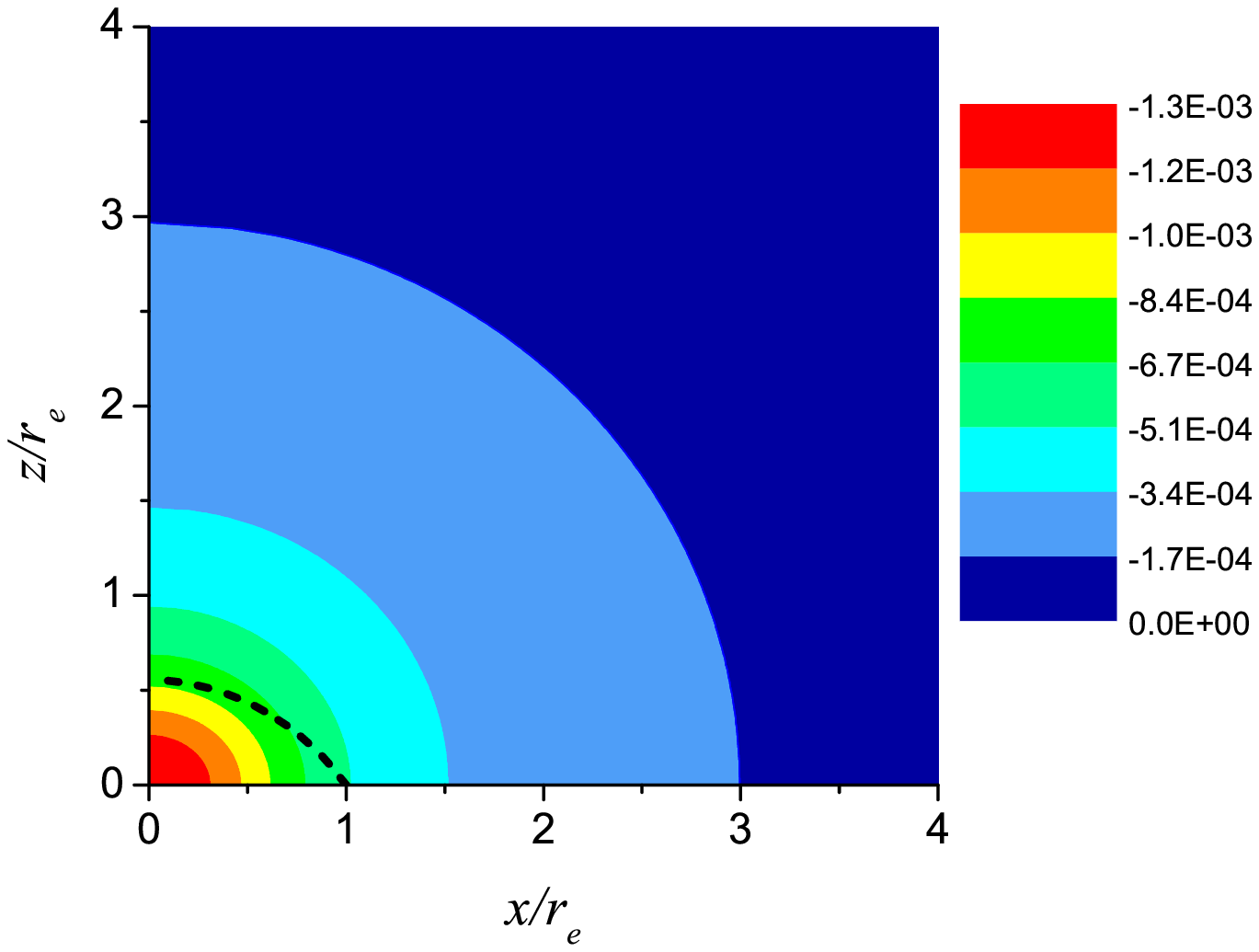}
\caption{Contour plot of the neutron star scalar field $\varphi$ in the
nonrotating case (left panel) and for a model rotating at the
mass-shedding limit (right panel); see text for details.
The thick, dashed line represents the neutron star surface.}
\label{Fig:Phi_contour_alpha}
\end{figure}

A contour plot of the scalar field is shown in
Fig. \ref{Fig:Phi_contour_alpha}. The left panel displays the scalar
field of a static neutron star, while the right panel shows
the scalar field of a star rotating at the mass-shedding limit.
 The Cartesian coordinates $(x,z)$ are obtained by the usual
transform of spherical polar coordinates and are normalized by
the equatorial coordinate radius $r_e$. The surface
of the star is shown as a thick dashed line.
The two models have the same central energy density
${\tilde \varepsilon}_c=1.25\times 10^{15} g/cm^3$ while their mass
is $M=1.77M_\odot$ and $M=2.20 M_\odot$, correspondingly.
In the rapidly rotating case, the isosurfaces of the energy density and
the scalar field no longer coincide -- the
flattening of the scalar field isosurfaces due to the rotation
is weaker than for the fluid variables, which is an expected effect.
In the right panel, a well known effect could also be observed --
the surface of the star forms a cusp at the equator, when reaching
the mass-shedding limit.

\subsection{Scalar-tensor theories with $\ln {\cal A}(\varphi) = \frac{1}{2} \beta \varphi^2$} \label{Sec:Results_SecondSTT}
This example of STT is of particular interest, because, it is
indistinguishable from GR in the weak field regime but it can differ
significantly when strong fields are considered. In the latter case,
all solutions of the GR field equations are also solutions of the
STT field equations with a trivial scalar field, but additional
solutions can also exist. For example, for certain values of the
 parameter $\beta$ and in a certain range of central densities
nonuniqueness of the neutron star solutions is present -- in
addition to the solutions describing neutron stars with a zero scalar
field (trivial solutions), neutron star solutions with a nonzero
scalar field exist (nontrivial solutions), which are also
energetically more favorable (spontaneous scalarization).

\begin{figure}[ht!]
\centering
\includegraphics[width=0.48\textwidth]{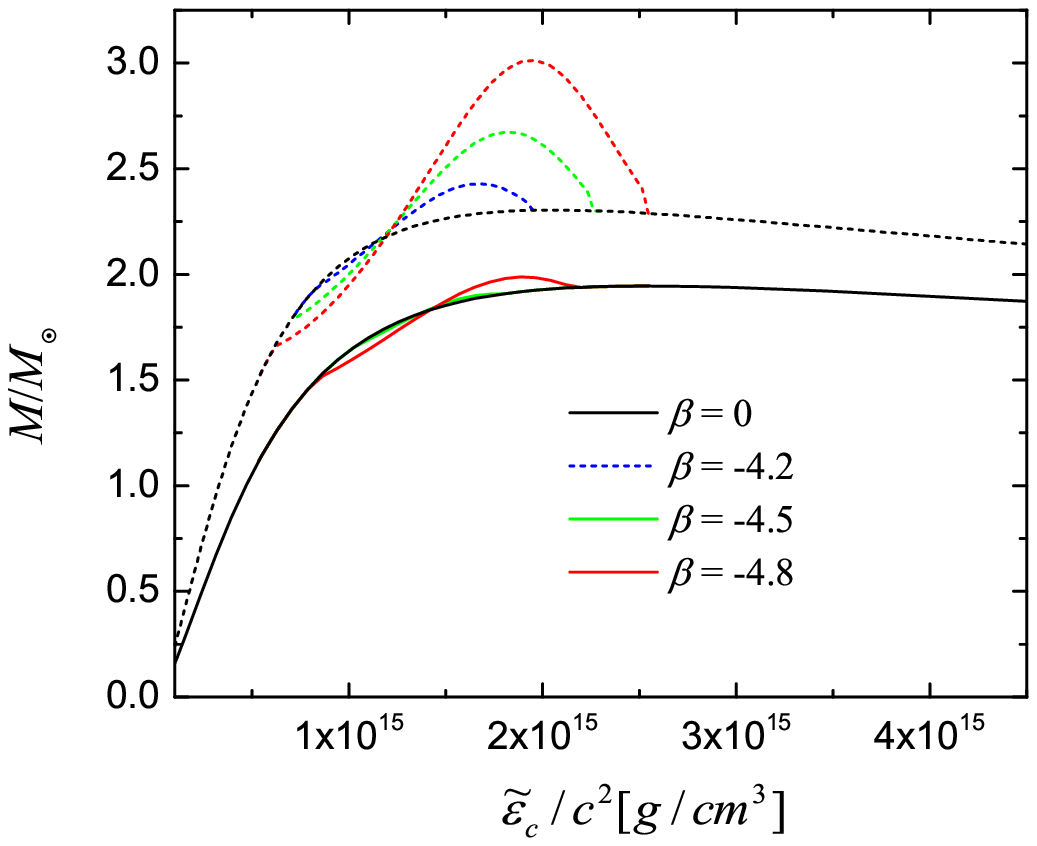}
\includegraphics[width=0.48\textwidth]{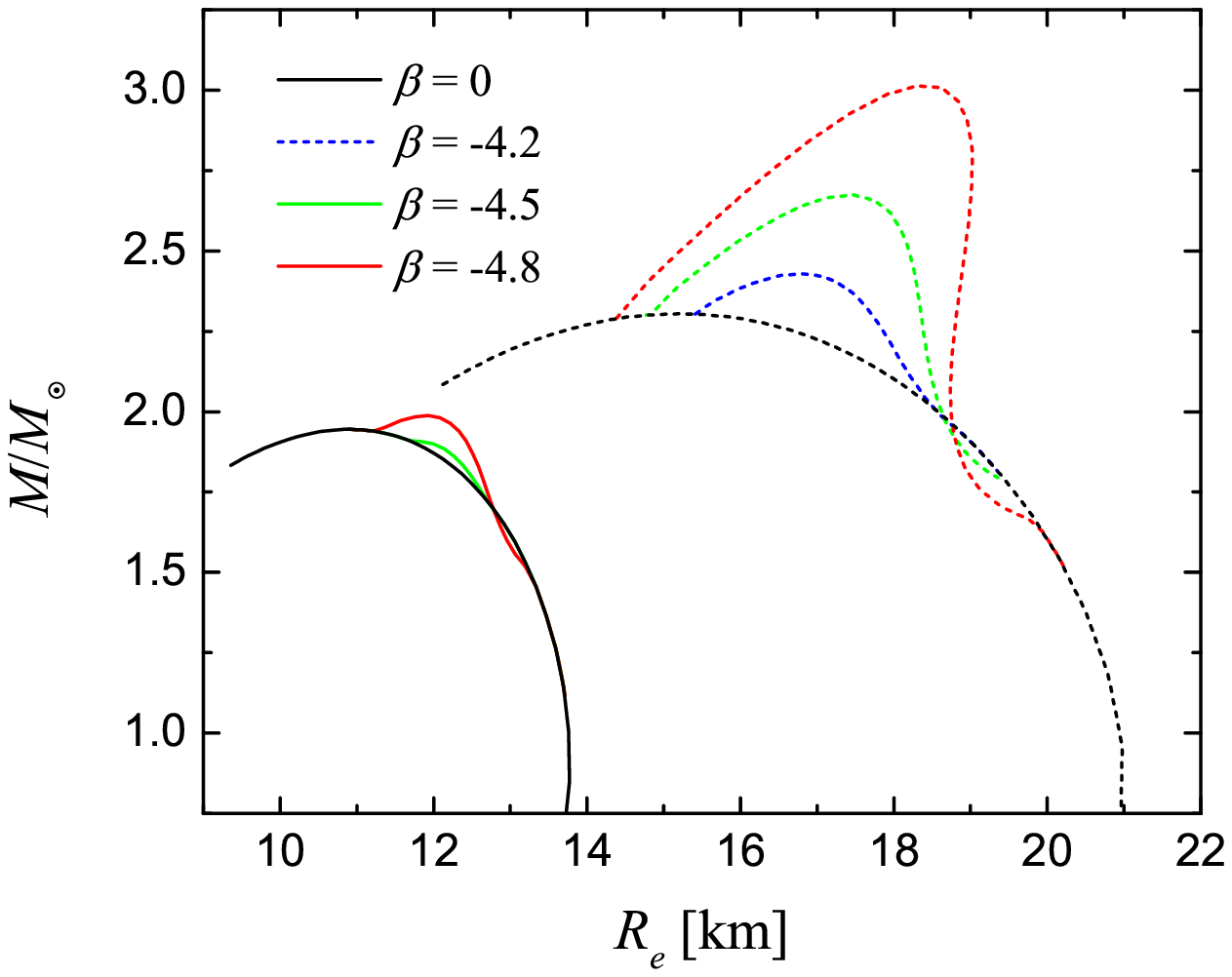}
\caption{The mass as a function of the central energy density
\emph{(left panel)} and of the radius \emph{(right panel)} for static
sequences of neutron stars (solid lines) and sequences of stars
rotating at the mass-shedding limit (dotted lines).
The trivial solutions coincide with the GR limit ($\beta=0$). For $\beta=-4.2$ nontrivial solutions do not exist in the nonrotating case. } \label{Fig:M_R_eps}
\end{figure}

We will consider three different values of the parameter
$\beta$: $-4.8$, $-4.5$ and $-4.2$.
The value $\beta=-4.8$ is the lower limit set by the binary
pulsar experiments. On the other hand, according to \cite{Harada1998}
spontaneous scalarization should only occur for $\beta<-4.35$ in the nonrotating
case. As an intermediate value, we choose $\beta=-4.5$.
Because rapid rotation allows for spontaneous scalarization
for larger values of $\beta$ than in the nonrotating case
(as we demonstrate below), we also
use the value of $\beta=-4.2$.

To investigate the effect of rapid rotation, we compare
solutions at the mass-shedding limit to nonrotating solutions, for
the chosen values of $\beta$. Fig. \ref{Fig:M_R_eps} shows
the mass as a function of the central energy density (left panel)
and as a function of the circumferential equatorial
radius. The black lines represent the solutions with zero scalar field,
and the colored lines correspond to different nonzero
values of $\beta$. Solid lines are nonrotating models,
while models at the mass-shedding limit are shown as dotted lines.
Equilibrium models with intermediate rotation rates exist in-between
the two limits. We observe that the branching of the neutron star
solutions with a nontrivial scalar field from the GR solutions occurs
for a larger range of central energy densities for rapidly rotating
stars than for nonrotating ones. In addition, we find that it occurs
for larger values of $\beta$ (such as -4.2), for which there are no
nontrivial solutions in the nonrotating case.

\begin{figure}[ht!]
\centering
\includegraphics[width=0.48\textwidth]{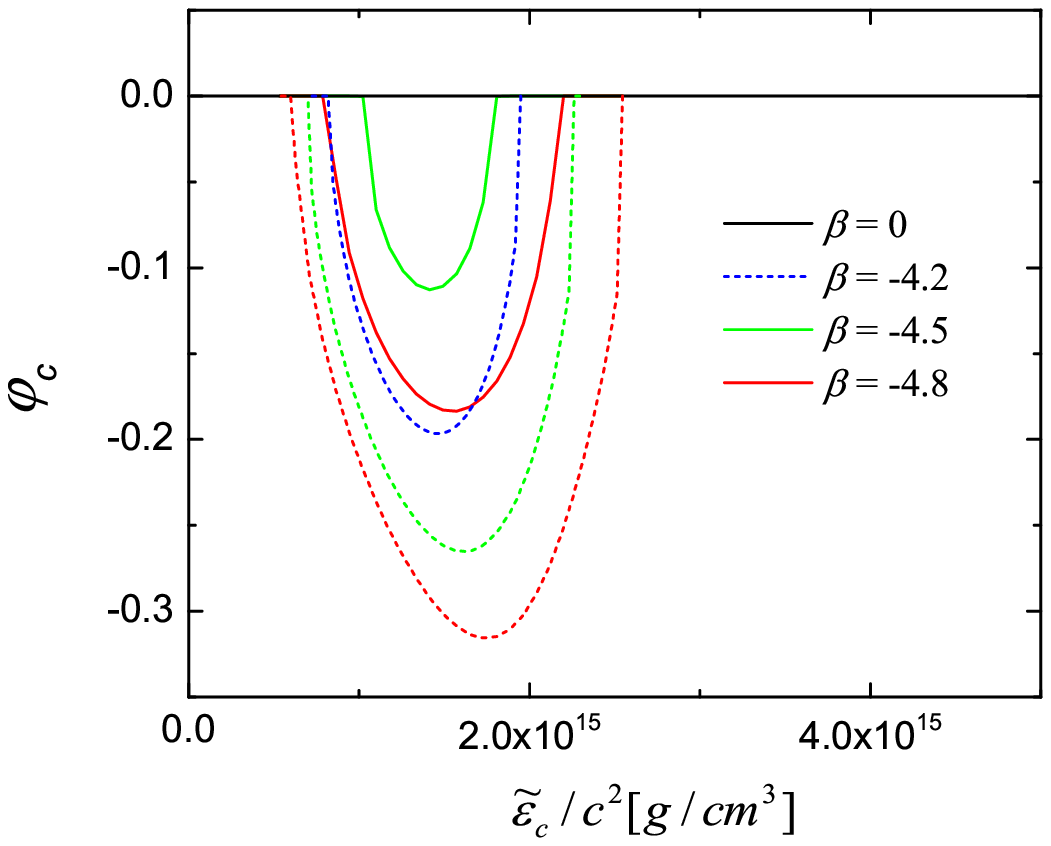}
\includegraphics[width=0.48\textwidth]{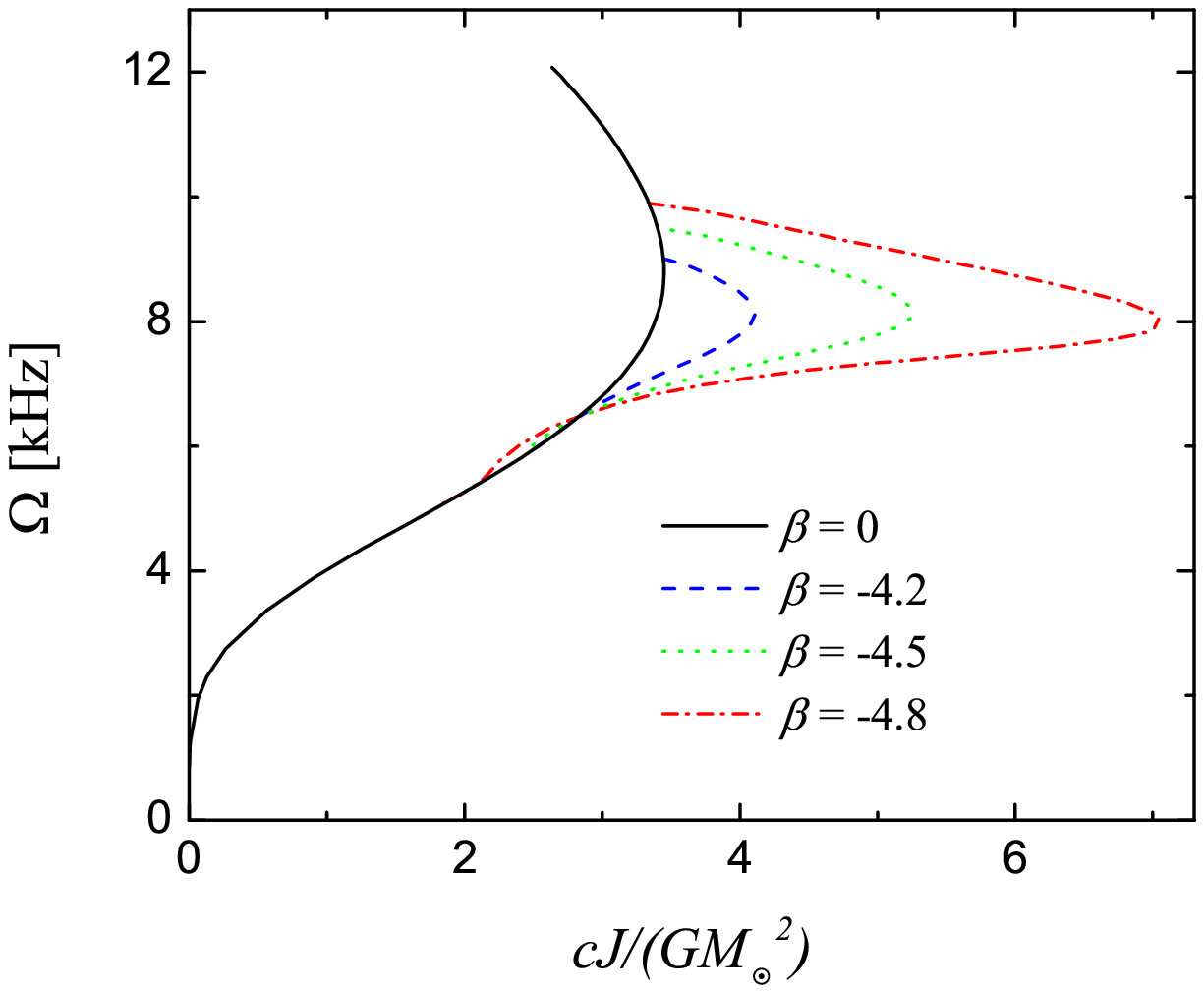}
\caption{\emph{(left panel)} The central value of the scalar field as a function of the central
energy density  for static
sequences of neutron stars (solid lines) and sequences of stars
rotating at the mass-shedding limit (dotted lines). For $\beta=-4.2$ nontrivial solutions do not exist in the nonrotating case. \emph{(right panel)} The angular velocity as
a function of the angular momentum for sequences of stars
rotating at the mass-shedding limit.}
\label{Fig:Om(J)phic_eosII.eps}
\end{figure}

Fig. \ref{Fig:Om(J)phic_eosII.eps} shows the central value of the scalar field as a function
of the central energy density (left panel)\footnote{The figure shows only the negative scalar field solutions -- as we mentioned earlier, a second branch of solutions exists with positive scalar field and with the same global characteristics, such as mass, radius, angular momentum, etc.} and the angular
velocity of the star as a function of the normalized
angular momentum (right panel). At a certain critical central energy density
$\varepsilon_{crit}^{min}$, the nontrivial solutions
branch out from the trivial ones. As the central energy density is increased,
the scalar field first increases, reaches a maximum and then decreases again,
until the nontrivial branch of solutions merges with the
trivial one at $\varepsilon_{crit}^{max}$. Thus, nontrivial
solutions exist only for densities between
$\varepsilon_{crit}^{min}$ and $\varepsilon_{crit}^{max}$
and these two values depend on the parameter $\beta$ and on the rotation rate.
Although we only focus on a representative EoS here, this range of central
energy densities is also EoS dependent.

The difference in mass,
radius and angular momentum between the trivial and the nontrivial
solutions increases with the rotation rate. For example,
the nontrivial models with $\beta=-4.5$ have almost the same
equilibrium properties as the trivial ones in the static limit,
but at rapid rotation the strength of the scalar field increases significantly
and then the mass, radius and angular momentum also differ considerably.
Moreover, for $\beta=-4.2$ there are no nontrivial solutions in
the static limit, but such solutions exist for fast rotation.
This means that the maximum value of $\beta$ for which scalarization
is possible increases considerably for high rotational rates.

\begin{figure}[ht!]
\centering
\includegraphics[width=0.6\textwidth]{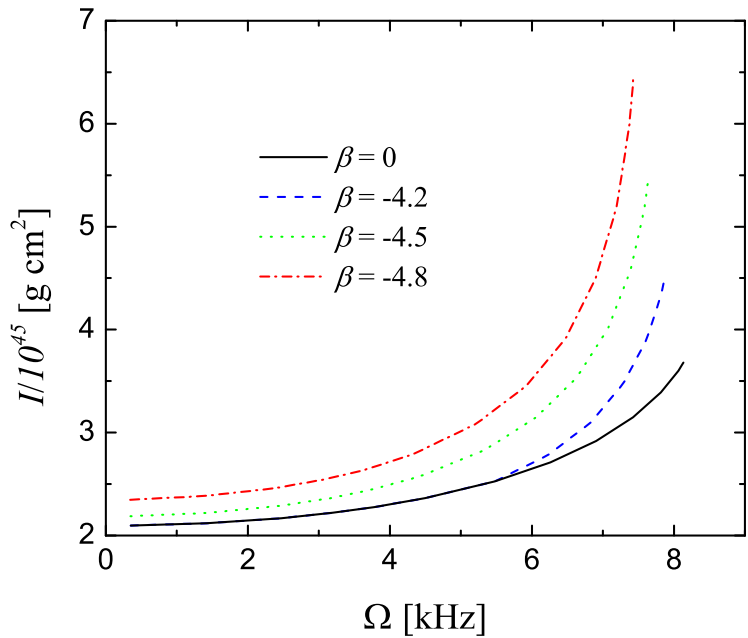}
\caption{Moment of inertia as a function of
the angular velocity, for different values of $\beta$.
All the models have the same central energy density
${\tilde \varepsilon}_c = 1.5 \times 10^{15}\; {\rm g/cm^3}$.}
\label{Fig:I(Om)_eosII.eps}
\end{figure}

Fig. \ref{Fig:I(Om)_eosII.eps} shows
the moment of inertia, $I=J/\Omega$, as a function of the
angular velocity, for sequences with a fixed value of the
central energy density. The moment of inertia can reach
higher values for scalarized models, compared to the trivial
solution, at all rotation rates (even at slow rotation)
for $\beta=-4.8$ and $\beta=-4.5$. In the case of $\beta=-4.2$,
nontrivial solutions do not exist in the nonrotating limit, but above
a certain rotational frequency the nontrivial branch of solutions appears.

From the above figures it is evident that the strongest effect of
 the scalar field for rotating models is on the angular momentum and
the moment of inertia of the star, which can differ up to a factor of
two, compared to the GR case, for the lowest possible value of
$\beta=-4.8$. Because the moment of inertia is affected even at slow rotation,
it could become useful as a sensitive astrophysical probe.
Other observables, such as the frequencies of emitted gravitational
waves (when stars are dynamically perturbed) should also be affected by
the presence of a nontrivial scalar field.

The distribution of the scalar field inside the star is
qualitatively very similar to the one shown in
Fig.~\ref{Fig:Phi_contour_alpha} for $\ln {\cal A}(\varphi) =  k_0 \varphi$, that is,
the rotation causes flattening  of the isosurfaces of the scalar field, which is weaker
than the flattening of the matter, so that the scalar field isosurfaces
do not coincide with the density isosurfaces. Instead, the density isosurfaces coincide
with the isosurfaces of an effective potential, that can be defined through
\eqref{eq:Hydrostatic_Equil}, requiring a barotropic equilibrium.

\begin{figure}[ht!]
\centering
\includegraphics[width=0.48\textwidth]{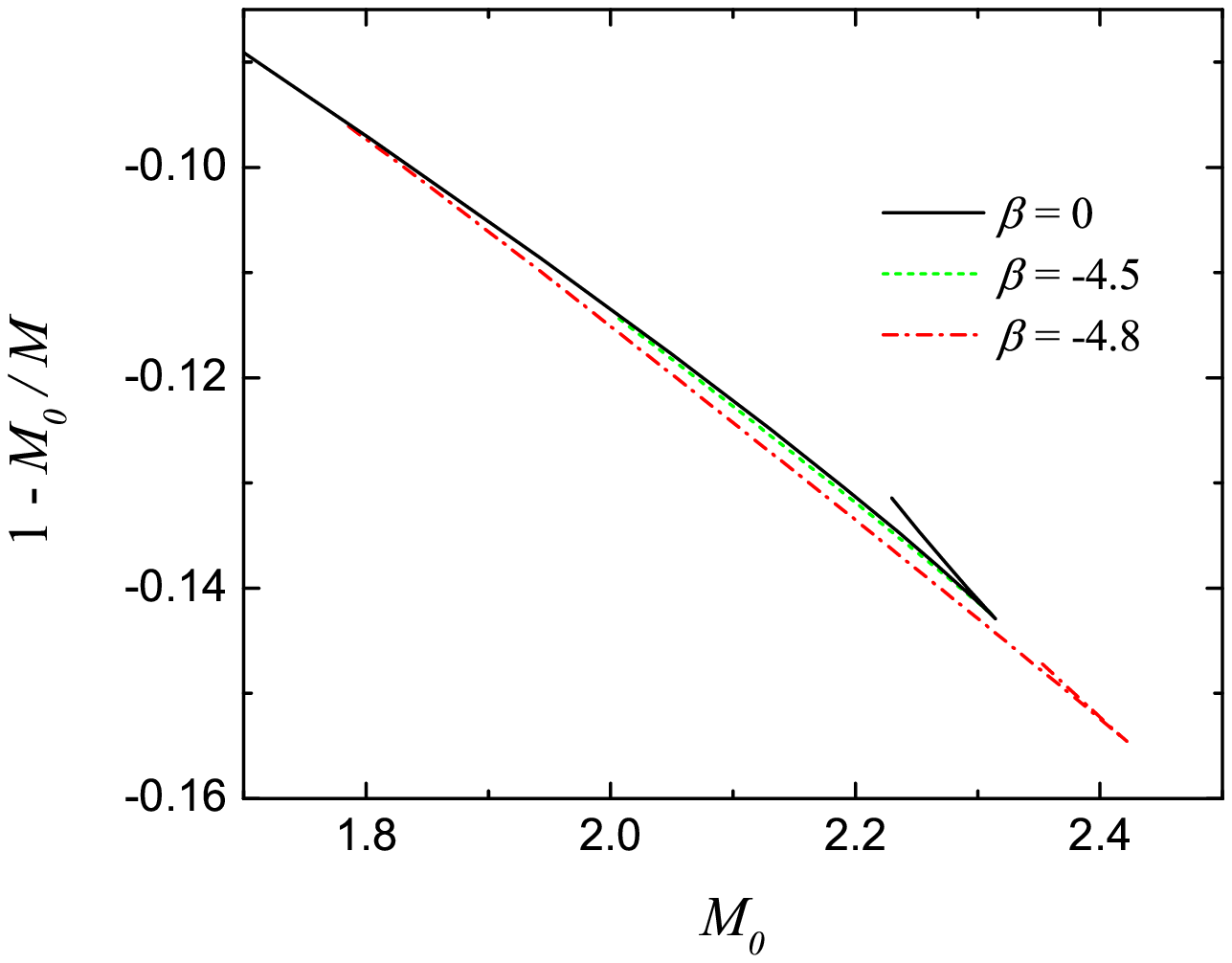}
\includegraphics[width=0.48\textwidth]{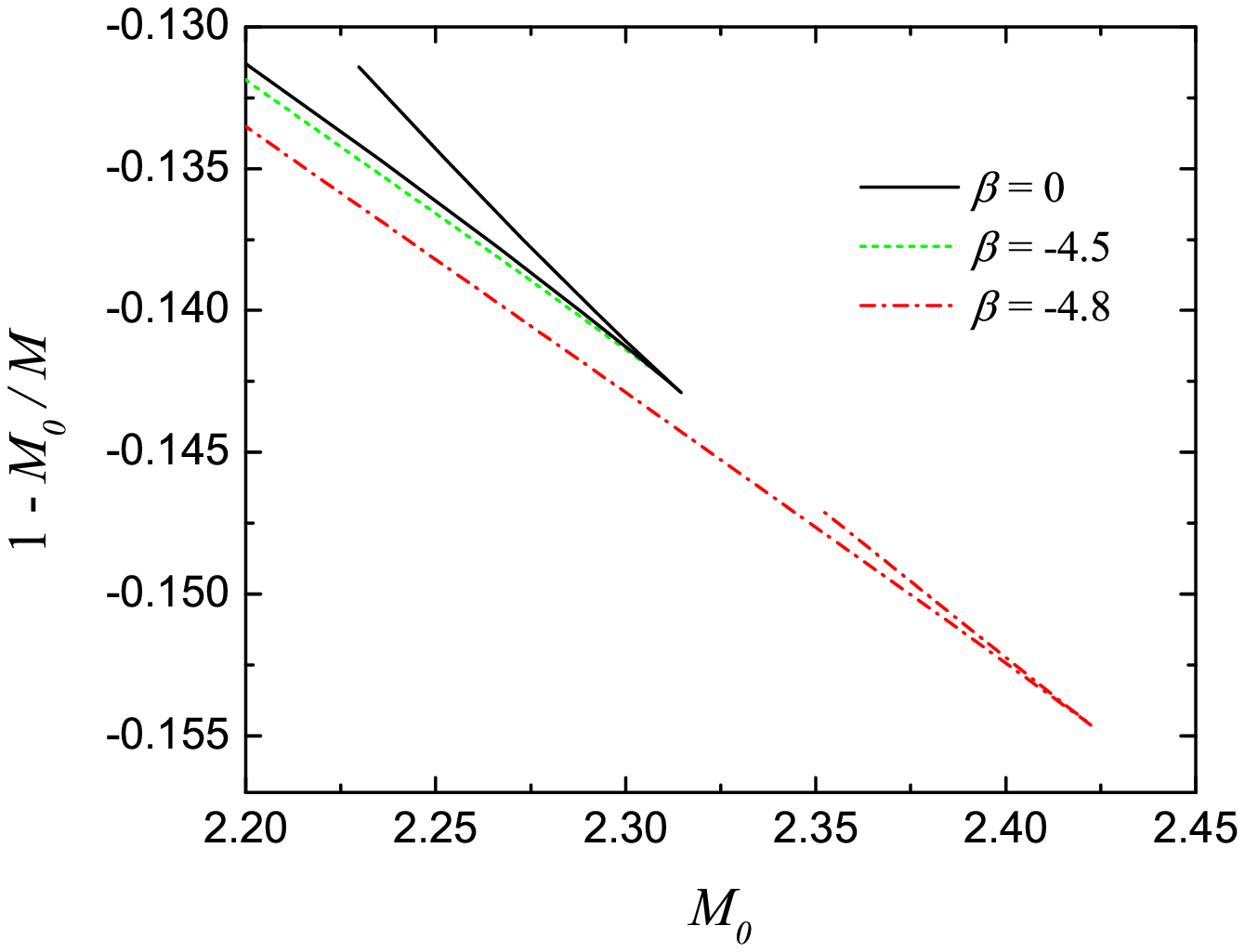}
\caption{The relative energy $1-M_0/M$ as a function of the baryon rest mass for a fixed value of the angular momentum $cJ/(GM_\odot^2)= 1.38$. In the right panel a magnification of the turning point area is shown.}
\label{Fig:BingingEn.eps}
\end{figure}

In the static case, it was shown that for a given baryon mass the
scalarized neutron stars have a lower total mass compared to the
trivial solution (when both solutions exist) and therefore
they are energetically more favorable \cite{Damour1993}.
In Fig. \ref{Fig:BingingEn.eps} we plot the {\it relative energy}
$1-M_0/M$, versus the rest mass for a given, representative
value of the angular momentum $cJ/(GM_\odot^2)= 1.38$.
We find that all nontrivial solutions have a lower relative energy and
are thus energetically more favorable than the corresponding trivial solutions.
This result holds also for any other value of the angular momentum
in the parameter space where both trivial and nontrivial solutions are possible.
 In Fig. \ref{Fig:BingingEn.eps}, each line folds over at a cusp, which
represents a turning point along the fixed-$J$ sequence, where secular
instability to collapse sets in. To make this clearer, we show a magnification
of the cusp area in the right panel of Fig. \ref{Fig:BingingEn.eps}.

As we pointed out in Section \ref{Sec:Results_SecondSTT}, nontrivial solutions can exist in the nonrotating case only
if $\beta < -4.35$ (for equation of state II \cite{Harada1998}). Determining a corresponding limit in the rapidly rotating
case requires some care in choosing appropriate initial conditions for the numerical iteration, when
considering central densities for which only the trivial solution exists in the nonrotating limit.
We find that along the sequence of models rotating at the mass-shedding limit, nontrivial solutions
can exist for approximately $\beta < -3.9$, which is a significantly less stringent limit than in the nonrotating case.
A detailed study of the critical $\beta$ at different rotation rates, as well as for different equations of state,
will be presented in a forthcoming publication.

\section{Summary and discussion}\label{Sec:Conclusions}
For the first time, we present the field equations governing rapidly
rotating neutron stars in scalar-tensor theories of gravity. We solve
these equations numerically, by extending the {\tt rns} code
\cite{Stergioulas95}, implementing the required modifications coming from
the scalar field. The accuracy of our numerical solutions was checked
by a comparison with an independent code in the slow-rotation approximation.

For these first, representative numerical solutions, we
employ a specific polytropic equation of state, that
is widely used in the literature on nonrotating neutron stars in STT.
Furthermore, we assume rigid rotation and a vanishing
potential $V(\varphi)$. We consider two examples of STT --
when $\ln {\cal A}(\varphi) =  k_0 \varphi$ and when
$\ln {\cal A}(\varphi) = \frac{1}{2} \beta \varphi^2$, where $k_0$
and $\beta$ are constants. In the first case,
which is equivalent to the Brans-Dicke theory, all of the
solutions possess a nontrivial scalar field, but $k_0$ is limited to
low values by observations, which leads
to a very weak scalar field and models that differ only
marginally from their GR counterparts. Our results
show that the strength of the scalar field could increase for rapid
rotation, but still the total mass, radius and angular momentum of
the STT models are almost indistinguishable from the GR
case, even for stars rotating at the mass-shedding limit.

The second example is more interesting, because it is perturbatively
equivalent to GR, but significant deviations from GR
are present when strong fields are considered. For
example, it was shown that in the static case, for certain values
of $\beta$, nonuniqueness of the neutron star solutions  is
present -- in addition to the neutron stars with a trivial scalar
field, additional nontrivial solutions appear, which are energetically
more favorable \cite{Damour1993}. We find that such nontrivial solutions are
present also in the rapidly rotating case for a larger range of central energy
densities. The critical value of $\beta$ for which scalarized neutron stars exist also changes from $\beta < -4.35$
in the nonrotating case to approximately $\beta < -3.9$ for models rotating at the mass-shedding
limit. Furthermore, we show that the nontrivial solutions are energetically
more favorable than the corresponding trivial ones.

For a given value of $\beta$, the changes (when compared to GR)
in the equilibrium properties of scalarized models,
such as mass, radius and angular momentum, increase with rotation,
which is due to an increased strength of the scalar field. For
example, for the limiting value of $\beta$ set by observations
($\beta=-4.8$), the angular momentum of neutron stars with
a nontrivial scalar field rotating at the mass-shedding limit can
be up to two times larger than for the corresponding models with a zero
scalar field.  It is interesting that the effect of scalarization on
the moment of inertia is significant even in the slow-rotation limit.
Such large changes in the neutron star properties
may lead to interesting observational effects.

Even though normal pulsars rotate at most at a small fraction of the
mass-shedding limit, it is conceivable that at least
magnetars are born rapidly rotating, in which case our results
would apply. The subsequent spin-down could cause a transition from
a scalarized solution to a trivial solution, as the nonrotating limit
or the axisymmetric instability limit are approached.
Furthermore, the recycling of old neutron stars through accretion takes the stars
through the region where the scalarized solutions are energetically more
favorable. Such scenarios are worth exploring for possible
observational signatures. In forthcoming publications, we are planning
a more detailed study of models with different equations of state and rotation laws,
as well as a study of the astrophysical implications of the scenarios
mentioned briefly above.

\section*{Acknowledgements}
We are grateful to Th. Apostolatos and G. Pappas for valuable comments.
DD would like to thank the Alexander von Humboldt Foundation for support. KK and SY would like to thank the Research Group Linkage Programme of the Alexander
von Humboldt Foundation for the support. The support by the Bulgarian National Science Fund
under Grant DMU-03/6, by the Sofia University Research Fund under Grant 33/2013 and by the German Science Foundation (DFG) via
SFB/TR7 is gratefully acknowledged. NS is grateful for the hospitality of the T\"ubingen
group during an extended visit.


\bibliography{references}

\end{document}